\documentclass[onecolumn,12pt]{article}
\pdfoutput=1

\usepackage{amssymb} 
\usepackage[sort&compress,numbers,square,comma]{natbib}  
\usepackage{times} 
\usepackage{amsmath}
\usepackage[affil-it]{authblk} 
\usepackage{simplewick}
\usepackage{fancyhdr}
\usepackage[pdftex]{graphicx} 

\newcommand{\beq}{\begin{equation}}
\newcommand{\eeq}{\end{equation}}
\newcommand{\ket}[1]{\left | #1 \right \rangle}                 
\newcommand{\bra}[1]{\left \langle #1 \right |}			
\newcommand{\balign}{\begin{equation}\begin{aligned}}
\newcommand{\ealign}{\end{aligned}\end{equation}}
\newcommand{\aph}{\alpha'}
\newcommand{\ah}{\alpha}
\renewcommand{\b}{\beta}
\renewcommand{\d}{\delta}
\newcommand{\D}{\Delta}
\newcommand{\ep}{\epsilon}
\newcommand{\g}{\gamma}
\newcommand{\G}{\Gamma}
\newcommand{\varep}{\varepsilon}
\renewcommand{\l}{\lambda}
\newcommand{\m}{\mu}
\newcommand{\n}{\nu}
\renewcommand{\o}{\omega}
\renewcommand{\k}{\kappa}
\renewcommand{\l}{\lambda}
\renewcommand{\r}{\rho}
\newcommand{\s}{\sigma}
\newcommand{\p}{\psi}
\newcommand{\pt}{\widetilde{\psi}}

\newcommand{\bz}{\bar{z}}
\newcommand{\bw}{\bar{w}}

\newcommand{\ztwo}{z_1-z_2}
\newcommand{\zthree}{z_1-\bar{z}_2}
\newcommand{\zfour}{\bar{z}_1-z_2}
\newcommand{\zfive}{\bar{z}_1-\bar{z}_2}
\newcommand{\zsix}{z_2-\bar{z}_2}
\newcommand{\corl}[1]{\left \langle #1 \right \rangle}
\renewcommand{\sl}{$SL(2,\mathbb{R})$}
\renewcommand{\part}{\partial}
\DeclareMathOperator{\tr}{Tr}

\begin{document}
\title{High energy string-brane scattering for massive states}
\author[1]{William Black}
\author[2]{Cristina Monni}
\affil[1]{Queen Mary University of London \\
	Mile End Road \\
	E1 4NS \\
	United Kingdom}
\affil[2]{Dipartimento di Fisica, Universit\`{a} di Cagliari and INFN, Sezione di Cagliari\\
Cittadella Universitaria, 09042 Monserrato, Italy}
\date{}
 


\renewcommand{\thefootnote}{\fnsymbol{footnote}}  	
\begin{titlepage}
\fancypagestyle{plain}{%
\fancyhf{}
\renewcommand{\headrulewidth}{0pt}
\rhead{QMUL-PH-11-10}} 

\maketitle
\begin{abstract}
String-brane interactions provide an ideal framework to study the dynamics of the massive states 
of the string spectrum in a non-trivial background. We present here an analysis of tree-level amplitudes 
for processes in which an NS-NS string state from the leading Regge trajectory scatters from a D-brane 
into another state from the leading Regge trajectory, in general of a different mass, at high energies 
and small scattering angles. This is done by using world-sheet OPE methods and effective vertex operators.
We find that this class of processes has a universal dependence on the energy of the projectile. We then compare
the result for these inelastic processes with that which one would obtain from the eikonal operator in a non-trivial test
of its ability to describe transitions between different string mass levels. The two are found to be in agreement.
\end{abstract}
\end{titlepage}
%
\section{Introduction}
%
In its current state string theory provides us with a perturbative description of quantum gravity. It is 
unclear whether this description is valid in all regimes, or what the fundamental degrees of freedom might 
be, yet it is clear that it can be put to good use in learning about physics at the Planck scale where we 
expect gravitational interactions to become significant. Such insight may help us understand processes 
such as the formation of black holes, and their absorption of matter. Furthermore, by examining 
high energy scattering one may test the mathematical fortitude of string theory -- its ability to yield 
unitary amplitudes and finite quantum corrections -- as well as probing its short-distance structure. 

Early investigations into the high energy behaviour of scattering amplitudes in string theory examined two 
principal regimes, that of fixed scattering angle \citep{gross87a,gross87b,gross88} and that of fixed 
momentum transfer \citep{amati87a,amati87b,amati88}. In the former it was discovered that at each order in
the perturbative expansion the high energy behaviour of all string scattering amplitudes is dominated by a 
saddle point in the moduli space of the corresponding Riemann surfaces. As a result one could interpret 
these processes in terms of a classical string trajectory in spacetime. However, in the limit for which 
these results are valid one encounters difficulties with the convergence of the perturbative expansion 
\citep{mende89} and application of these
results is potentially problematic. The latter investigations consider graviton-graviton scattering in the 
limit of large centre of mass energy and fixed momentum transfer, referred to as the Regge limit; these 
revealed some interesting features arising at different values of the transferred momentum $t$. For small 
$t$, corresponding to large distance processes, it was found that elastic scattering dominates and this is 
mediated by the long range exchange of gravitons. For larger values of $t$ one encounters a semiclassical
eikonal description which reveals effects attributable to classical gravity, that is to say, the strings
begin to feel the influence of an effective curved background. If the value of $t$ is increased further 
absorptive processes begin to contribute significantly due to the production of inelastic and diffractive 
states and beyond this the analysis can be extended into the regime of small but fixed scattering angle.

More recently there has been renewed interest in the high energy behaviour of scattering amplitudes in 
string theory \citep{veneziano04,amati07,d'appollonio10}. In \citep{d'appollonio10} the methods 
employed in \citep{amati87b} are used to examine the high energy scattering of a graviton from a stack of 
$N$ parallel D$p$-branes in a flat background. Significantly, this process involves the interaction 
between a perturbative and non-perturbative object for which we have a microscopic description, whereas 
previously only interactions between perturbative states had been investigated. As in the earlier 
investigations the limit of large energies is examined in a number of kinematic regions for which both
the impact parameter and scale of curvature\footnote{by which we mean $R_p$, see 
equation~(\ref{eq:curvature})} produced by the D-branes are larger than the string length. 
For the largest impact parameters one finds a region of perturbative scattering leading to vanishingly
small deflection angles, and as the impact parameter is reduced one encounters both classical corrections
and corrections due to the nonzero string length. As the impact parameter approaches the scale of 
curvature one expects to see gravitational effects and the string can be considered classically to have 
been captured; it has been suggested that this would be an interesting region for further study, together
with impact parameters much smaller than the curvature, since an understanding of these regions could provide
one with the tools required to study the microscopic process of matter falling past a horizon.

In this paper we are motivated by these previous works to study the interactions of a massive closed 
superstring with a stack of D$p$-branes. Such interactions will in general include inelastic processes in 
which the string is excited/decays into a state of a different mass. Since we expect scattering at large 
energies to be particularly simple for the leading Regge trajectory -- those states in the string spectrum with the 
maximum spin possible for their mass -- we will take such states as our two external strings. The presence
of the D-branes will break the $SO(1,9)$ invariance to $SO(1,p)\times SO(9-p)$ and momentum will only be 
conserved parallel to the world volume of the D-branes; as such, the invariant quantities we will 
primarily consider here shall be composed of the momentum flowing parallel to the D-branes 
and the momentum transferred to the D-branes and after a suitable transformation the former becomes the energy
of the projectile. We will be interested in the limit in which this energy becomes extremely large in comparison 
to the string scale while the magnitude of the momentum exchanged is kept fixed. In order to obtain high energy 
scattering amplitudes in the most efficient manner we will make use of operator product expansion (OPE) methods to construct effective
vertex operators as pioneered in \citep{ademollo89,ademollo90} and more recently used in 
\citep{brower06,fotopoulos10}. Insertion of these effective vertices onto the upper-half of the complex
plane will yield the tree-level amplitudes of these processes which show identical Regge behaviour to 
that exhibited in the graviton-graviton scattering. This behaviour is universal for the states on 
the leading Regge trajectory. An analysis using the eikonal operator of \citep{d'appollonio10} gives
results in agreement with those produced by world sheet calculations, providing further verification of 
its capability in computing elements of the string S-matrix. This is a non-trivial confirmation of the 
efficacy of the eikonal operator in analysing scattering beyond the case of massless states. Furthermore,
these checks demonstrate by the explicit calculation of string amplitudes that longitudinal excitations
are absent in the scattering of a massive string with a D-brane for large impact parameters, as indicated
in \citep{d'appollonio10} by computations concerning the quantisation of a string in the background
generated by the D-branes.

The contents of this paper are as follows. In Section~\ref{sec:kinematics} we lay out our conventions for
the kinematics of string-brane scattering and describe the particular regions of the parameter space in 
which we are interested. In Section~\ref{sec:spectrum} we introduce the string states 
that we will use, that is, the states of the NS-NS sector in type II string theory with maximal spin in 
comparison to other states with the same mass; these states form the leading Regge trajectory. As an 
example it is shown how to determine the BRST invariant vertex operator for the first massive state on the 
leading Regge trajectory, and this result is then generalised to give the vertex operator for a state with 
mass $\aph M^2=4n$. In Section~\ref{sec:amp_comp} we consider the amplitude for the inelastic excitation
of a state with $n=0$ to a state with $n'=1$ and study its high energy limit. This direct evaluation of 
the amplitudes can then be compared to the results of Section~\ref{sec:ope_methods} in which we construct 
an effective vertex using OPE methods in order to compute the leading high energy contribution for a
generic inelastic process involving states of the leading Regge trajectory. In Section~\ref{sec:eikonal} 
we show that in the Regge limit the tree-level string amplitudes are precisely reproduced by the matrix
elements of the eikonal operator introduced in \citep{d'appollonio10} and hence briefly examine what this
may imply for the scattering of massive states at large impact parameters. In Section~\ref{sec:conclusion} 
we discuss these results and their implications for other high energy processes. In particular we comment
on the interesting limit of very large masses for the external states.
%
\section{Kinematics of a string scattering from a D$p$-brane}
\label{sec:kinematics}
%
Here we briefly review the kinematics appropriate to tree-level interactions between a single string and
a D$p$-brane. Throughout this work we will use a flat metric of positive signature, $\eta=diag(-,+,+,...,+)$. 
In the following computations we will consider some initial state with momentum $p_1$ such that 
$\aph M_1^2=-\aph p_1^2=4n$; after interacting with the D$p$-brane we are left with another state with 
momentum $p_2$ which in general can have a different mass, $\aph M_2^2=-\aph p_2^2=4n'$. We can decompose
these momenta into vectors which are parallel and orthogonal to the directions in which the D-brane is 
extended, that is
\beq
	p_i^{\m}=p_{i \Vert}^{\m}+p_{i \bot}^{\m}.
\eeq	
The mass of a D-brane scales as $1/g_s$ and so to leading order in the perturbative expansion it is 
infinitely massive. Consequently we may neglect its recoil and momentum is only conserved in directions 
parallel to the D-brane,
\beq
	p_{1 \Vert}^{\m}+p_{2 \Vert}^{\m}=0.
\eeq
For brevity we denote $p_{1 \Vert}^{\m}=-p_{2\Vert}^{\m}=p_{\Vert}^{\m}$. Using the mass-shell condition 
we may relate this quantity to the orthogonal components of momentum
\begin{subequations}
\begin{align}
	p_{1\bot}^2=&-p_{\Vert}^2-\frac{4n}{\aph}, \\
	p_{2\bot}^2=&-p_{\Vert}^2-\frac{4n'}{\aph}. 
\end{align}
\end{subequations}
The vectors $p_{i\bot}$ are by definition space-like\footnote{ignoring the specific case of a D-instanton.}
so we can infer that $p_{\Vert}$ is necessarily time-like. Considering this we define the following 
kinematic invariants for use as the parameters in our computations
\begin{subequations}
\begin{gather}
	s\equiv-p_{\Vert}^2, \\
	t\equiv-(p_1+p_2)^2, \\
	E\equiv \sqrt{s}.
\end{gather}
\end{subequations}
After a Lorentz transformation to a frame in which the spatial components of $p_1$ are nonzero only
in the orthogonal directions the quantity $E$ will be equal to the energy of the string.

By definition, $s$ and $t$ have the following physical boundaries,
\begin{align} 
	\max\{M_1^2,\,M_2^2\}\le &s <\infty,\\  
\left(\sqrt{s-M_1^2} - \sqrt{s-M_2^2}\right)^2 \le |&t| \le
\left(\sqrt{s-M_1^2}+\sqrt{s-M_2^2}\right)^2.
\end{align}
We could explore various kinematic regimes, depending on the relative sizes of the initial and final mass 
of the string, the square of the string energy $s$ and the momentum transfer $t$. In this paper we will 
analyse the Regge regime of such amplitudes, meaning their behaviour for $\aph s\rightarrow\infty$ and 
$\aph t$ fixed. We will see that for massive string states from the leading Regge trajectory these 
amplitudes demonstrate typical Regge behaviour. More precisely, we will keep the external states fixed 
while taking the large $s$ limit, which implies that the kinetic energy of the projectile is very large.

As we will show in section \ref{sec:ope_methods}, if we consider Regge kinematics the leading contribution 
to the amplitudes is captured by the OPE of the vertex operators inserted far from the boundary of the 
world-sheet. The amplitudes could be dominated by the OPE and show Regge behaviour at high energy also 
when we consider external states with very large masses, as long as the difference $n'-n$ is negligible 
compared to $\aph s$ so that $t/s$ remains small. If instead we allow a large mass gap between initial and 
final states and we let it grow with the projectile energy,  $\Delta M\sim E$, then we are in a kinematic 
regime where $t\sim s$. In this regime we expect the leading contribution to the amplitude to be given not 
by the OPE but by a semiclassical world-sheet corresponding to a saddle-point in the moduli space. The 
amplitudes should then show an exponential decay in the energy, typical of scattering processes at fixed 
angle  \cite{gross87a, gross87b, gross88, gross89}. The analysis of the scattering amplitudes in the limit 
of large masses for the external states is however beyond the scope of this paper.
%
\section{Massive spectrum of the NS-NS superstring on the leading Regge trajectory}
\label{sec:spectrum}
%
In this section we analyze the massive spectrum of the NS-NS sector of the type II superstring, focusing 
on the highest spin states at a given mass, i.e, the leading Regge trajectory. As an illustrative 
example, we will review the BRST quantization of the first level, following the conventions used in 
\cite{polchinski98a,polchinski98b}. 

The mass shell condition for the $n^{th}$ level of the NS-NS sector of the superstring after the GSO 
projection is given by
\begin{equation}
 \alpha'M^2=4n,
\label{mass_shell}
\end{equation} 
with  $n=0,1,2...$. In the following we will write the closed string state as the product of two copies of 
the open string sector, so the states we will consider carry momentum $p^\mu=2k^\mu$ where $k$ will
represent the momentum of the open string vertex. Henceforth, when a result is stated for holomorphic 
fields, it is with the understanding that an analogous result will hold for the antiholomorphic 
quantities. 

For the first massive level the mass shell condition (\ref{mass_shell}) with $n=1$ is satisfied by the 
following states in the -1 picture:
\begin{subequations}
\begin{align}
	\ket{\phi_{\varep}}&=\varep_{\m \n}\alpha_{-1}^{\m}\p_{-\frac{1}{2}}^{\n}\ket{0;k}, \\
  	\ket{\phi_{A}}&=\mathcal A_{\m\n\r}\p_{-\frac{1}{2}}^{\m}\p_{-\frac{1}{2}}^{\n}\p_{-\frac{1}{2}}^{\r}\ket{0;k}, \\
	\ket{\phi_{B}}&=B_\m\p_{-\frac{3}{2}}^{\m}\ket{0;k}.
\end{align}
\end{subequations}
To find the physical states, we write the most general state at this level as a linear combination:
\begin{equation}
 	\ket{\phi}=\left( \varep_{\m\n}\ah_{-1}^{\m}\p_{-\frac{1}{2}}^{\nu}+\mathcal A_{\m\n\r}
		\p^\m_{-\frac{1}{2}}\p^\n_{-\frac{1}{2}}\p^\r_{-\frac{1}{2}}+B_\n\p^\n_{-\frac{3}{2}}\right) \ket{0;k}.
\label{general} 
\end{equation} 
The corresponding vertex operator with superghost charge $-1$ is
\begin{equation}
 V_{-1}=ce^{-\varphi}\left( \dfrac{i}{\sqrt{2\alpha'}}\mathcal \varepsilon_{\mu\nu}\partial X^{\mu}\psi^{\nu}+\mathcal A_{\mu\nu\rho}\psi^\mu\psi^\nu\psi^\rho+B_\nu\partial\psi^\nu\right) e^{ik\cdot X},
\label{vertex_1} 
\end{equation} 
where the field $c$ is one of the reparametrisation ghosts and the field $\varphi$, together with the 
$\eta$ and $\xi$ fields gives the bosonisation of the superghosts. In order for this state to be physical 
it must be invariant under BRST transformations.

As shown in \citep{bianchi10}, we can gauge away the scalar, the antisymmetric rank-2 tensor and the 
vector. The requirement of BRST invariance for the remaining states implies the constraints
\begin{equation}
	 k^{\mu}\mathcal \epsilon_{(\mu\nu)}=0, \quad k^\mu\mathcal A_{[\mu\nu\rho]}=0.
\label{phys_cond}
\end{equation}
We are left with two physical states:
\begin{itemize}
 	\item The state $\varep_{(\mu\nu)}\alpha_{-1}^{\mu}\psi_{-\frac{1}{2}}^{\nu}\ket{0;0}$ has a 
	polarization $\varep_{(\mu\nu)}$ which is a completely symmetric, traceless tensor invariant 
	under the litte group $SO(9)$, so it carries 44 degrees of freedom.
	\item The state $\mathcal A_{[\mu\nu\rho]}\p_{-\frac{1}{2}}^\m\p_{-\frac{1}{2}}^\n\p_{-\frac{1}{2}}^\rho$ has a polarization $\mathcal 
	A_{[\mu\nu\rho]}$ which is a three-form of $SO(9)$, corresponding to 84 degrees of freedom. 
\end{itemize}

Together, these two states have 128 degrees of freedom, which is the full bosonic content of the left 
sector of the first massive level, as explained in \cite{green87a}. 

As we consider higher and higher levels, it can be a difficult task to find the BRST invariant vertex 
operators, since the number of states increases dramatically. Nevertheless, at each level one could 
consider the states in the light-cone gauge, and these can always be rearranged into irreducible 
representations of the little group of $SO(9)$, which describe the BRST-invariant states.
This has been shown in \cite{bianchi10} for the first two levels. 

\begin{figure}
	\begin{center}
	\includegraphics[scale=0.6]{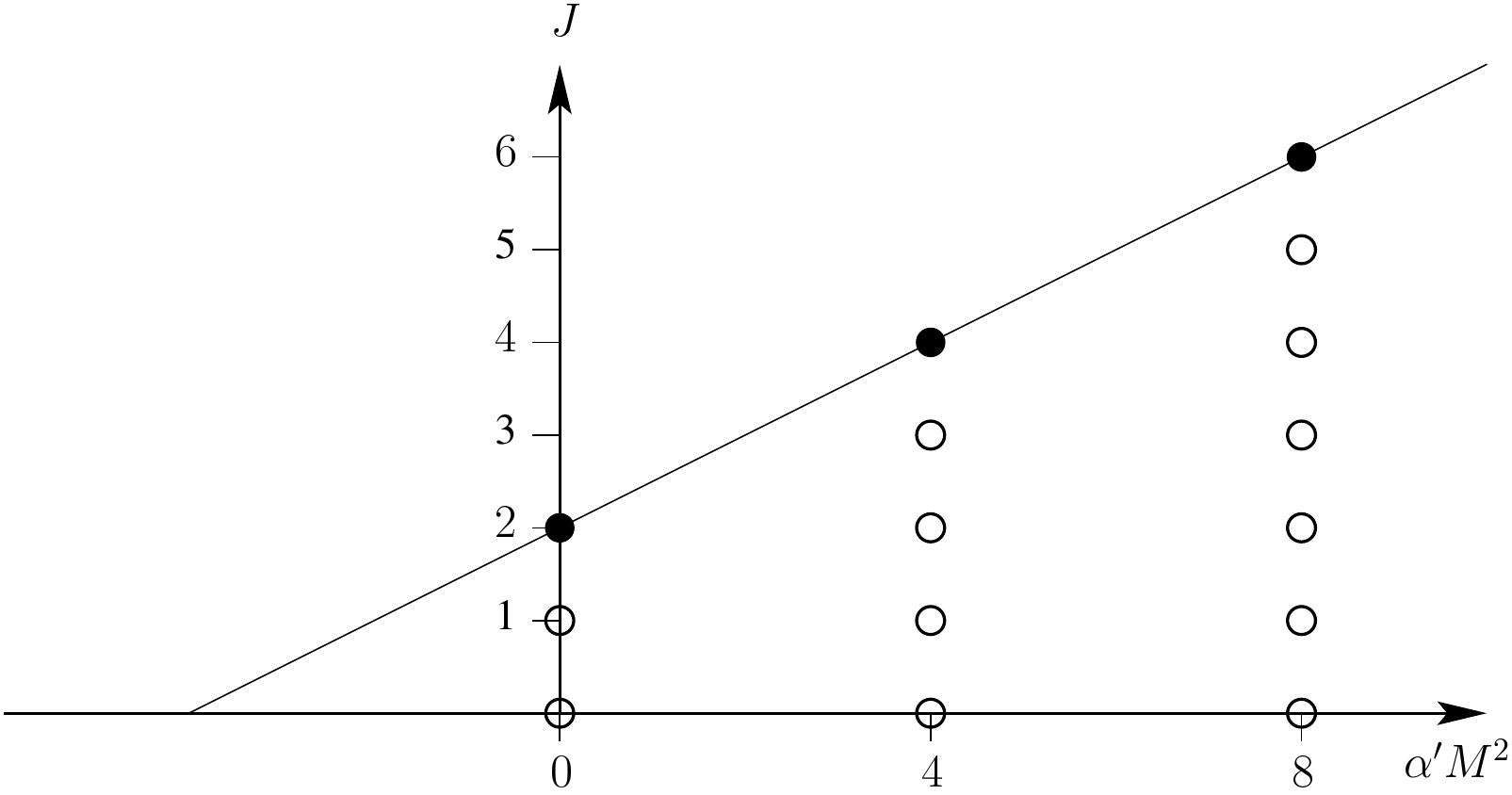}
	\end{center}
	\caption{Diagrammatic depiction of the type II string spectrum indicating the mass and spin of
		each physical state, represented by circles. Filled circles show states on the leading
		Regge trajectory.}
	\label{fig:regge_traj}
\end{figure}

In this paper we shall focus on states belonging to the leading Regge trajectory. The corresponding vertex 
operators have a particularly simple form. We first construct the vertex for an open string state which is 
totally symmetric in its polarisation. In general the tensor product of two such states will give a closed 
string state in a reducible representation containing physical states of the same mass but different spin. 
However, by symmetrising over all indices of the polarisation one is left with a state of spin 
$J=(\aph M^2+4)/2$. This is the maximum possible spin in the type II string spectrum for a fixed mass level 
and it is the set of all such states that comprises the leading Regge trajectory. This can be seen 
pictorially in figure~\ref{fig:regge_traj}. Written in terms of oscillators, such an open string state will 
have the form $\ket{\phi_n}=\varep_{\m_1\ldots \m_n\ah}\prod_{i=1}^n\ah_{-1}^{\m_i}\p_{-\frac{1}{2}}^{\ah}\ket{0;k}$, 
and this will give the following closed string state,
\beq 
	\ket{\phi_n}^{closed}= \epsilon_{\m_1\ldots \m_n\,\n_1\ldots \n_n(\ah\b)}\left[\prod_{i=1}^n\alpha_{-1}^{\m_i}
	\widetilde{\alpha}_{-1}^{\n_i}\right]\p_{-\tfrac{1}{2}}^{\ah}\widetilde{\psi}_{-\tfrac{1}{2}}^{\b}\ket{0;p},
\eeq
where $\epsilon_{\m_1\ldots \m_n \ah \n_1\ldots n_n\b}=\varep_{\m_1\ldots\m_n\ah} \otimes \widetilde{\varep}_{
\n_1\ldots\n_n\b}$. From here on in it shall be implicitly understood that the polarisation $\ep$ is symmetric in all indices.

By virtue of the state-operator isomorphism we obtain from the state above a vertex operator of superghost
charge (-1,-1),
\begin{equation}
	W_{(-1,-1)}^{(n)}(k,z,\bar{z})=
		\epsilon_{\mu_1\ldots\mu_{n} \alpha\,\nu_1\ldots \nu_{n} \beta}
		V^{\mu_1\ldots\mu_{n}\alpha}_{-1}(k,z)\widetilde{V}_{-1}^{\nu_1\ldots \nu_{n}\beta}(k,\bar{z})
	\label{eq:vertex_2},
\end{equation} 
with
\begin{equation}
	V^{\mu_1\ldots\mu_{n}\alpha}_{-1}(k,z)=\frac{1}{\sqrt{n!}}\left(\frac{i}{\sqrt{2\alpha'}}\right)^{n}e^{-\varphi(z)}
 		\left (\prod_{i=1}^{n}\partial X^{\m_i}\right )\psi^{\ah}e^{ik\cdot X(z)}.
\label{eq:vertex_2.5} 
\end{equation} 
Above we have written $X(z)$ for the open string field, obtained by decomposing the complete string 
coordinate field into $X(z,\bz)=(X(z)+\widetilde{X}(\bz))/2$, and as previously stated we make use of
the open string momentum $k$ for convenience while the physical momentum of the string is $p=2k$. Since
we wish to compute amplitudes on a world-sheet with the topology of the disc the vertices which make
up this amplitude should have a total superghost charge of $-2$, consequently we will need the vertex
operators of the state above in the $(0, 0)$ picture,
\begin{equation}
	 W_{(0,0)}^{(n)}(k,z,\bar{z})=-
		\epsilon_{\mu_1\ldots\mu_{n} \alpha\:\nu_1\ldots \nu_{n} \beta}
		V^{\mu_1\ldots\mu_{n}\alpha}_0(k,z)\widetilde{V}_0^{\nu_1\ldots \nu_{n}\beta}(k,\bar{z})
	\label{eq:vertex_1}
\end{equation} 
where
\begin{align}\nonumber
	V_0^{\mu_1\ldots \mu_{n}\alpha}(k,z)=\frac{1}{\sqrt{n!}}\left(\frac{i}{\sqrt{2\alpha'}}\right)^{n+1}\left(\partial X^{\mu_{n}}\partial X^{\alpha}
		-2i\alpha' k\cdot \psi \partial X^{\mu_{n}}\psi^{\alpha}-2\alpha' n\partial \psi^{\mu_{n}} \psi^{\alpha}\right)
\\ 
		\left (\prod_{i=1}^{n-1}\partial X^{\mu_i}\right )
		e^{ik\cdot X(z)}.
\label{eq:vertex_1.5}
\end{align} 
Before moving on we should make one point; strictly speaking, the vertices given in 
equations~(\ref{eq:vertex_1.5}) and (\ref{eq:vertex_2.5}) can only be applied to the cases $n>0$, 
however we shall extend their use to the case $n=0$ with the understanding that when one encounters 
a product of the form $\prod_{i=1}^0 a_i$ we are to replace it by unity.
%
\section{Computation of the amplitudes on the disk in the high energy limit}
\label{sec:amp_comp}
%
The scattering amplitude for the interaction of two closed string states with a stack of $N$ D$p$-branes 
at tree level is given by the insertion of two closed string vertex operators onto the upper-half of the 
complex plane,
\beq
	A_{n,n'}=\mathcal N \int_{\mathbb{H}_+}\frac{dz_1^2 dz_2^2}{V_{CKG}}\corl{W^{(n)}_{(0,0)}
		(k_1,z_1,\bz_1)	W^{(n')}_{(-1,-1)}(k_2,z_2,\bz_2)}_{\mathbb{H}_+}.
\label{eq:formal_amp}
\eeq
Here we have the vertex operator~(\ref{eq:vertex_1}) for a state with momentum $p_1=2k_1$ and mass 
$M^2=4n/\aph$ carrying a superghost charge $(0,0)$ and the vertex operator~(\ref{eq:vertex_2}) for a 
state with momentum $p_2=2k_2$ and mass $M^2=4n'/\aph$ carrying a superghost charge $(-1,-1)$.
In this section we will compute the amplitude involving one graviton and one massive symmetric state at 
the level $n'=1$. In doing so we shall see many features which are not only common to the methods 
introduced in Section \ref{sec:ope_methods} but also motivate them.

The normalization constant, $\mathcal N$, in (\ref{eq:formal_amp}) is formed from the product of the normalizations 
of the vertices, $\left(\k/2\pi\right)^2$,  and the topological factor for a disc amplitude $C_{D_2}= 2\pi^2
\:T_p/ \k$,  where $\k$ is the gravitational coupling constant in ten dimensions and $T_p$ is 
the coupling for closed string states to a D$p$-brane. Overall this gives the normalization
\beq
	\mathcal N=\dfrac{\kappa T_p}{2}
	=\dfrac{R_p^{7-p}\,\pi^{\frac{9-p}{2}}}{\Gamma\left( \frac{7-p}{2}\right) },
\label{eq:normalisation}
\eeq 
where $R_p$ represents a characteristic size for the stack of D-branes and is related to the t'Hooft 
coupling, $\l=gN$ as follows,
\begin{equation}
 	R_p^{7-p}=gN\dfrac{(2\pi\sqrt{\alpha'})^{7-p}}{(7-p)\Omega_{8-p}}, \qquad
 	\Omega_n=\dfrac{2\pi^{\frac{n+1}{2}}}{\Gamma\left( \frac{n+1}{2}\right) }.
	\label{eq:curvature}
\end{equation}  
Hence, using equations~(\ref{eq:vertex_2.5}--\ref{eq:vertex_1}) it is easily verified that the vertex for the 
graviton in the $(0, 0)$ picture is
\begin{equation}
	W_{(0,0)}^{(0)}(k_1,z_{1},\bar{z_{1}})=-\ep_{\mu\nu}V_{0}^{\mu}(k_{1},z_{1})
		\widetilde{V}_{0}^{\nu}(k_{1},\bar{z}_{1})
\label{eq:graviton_vertex_amp}
\end{equation} 
with
\begin{equation}
	V_{0}^{\mu}(k_{1},z_{1})= \dfrac{1}{\sqrt{2\alpha'}}\left(i\partial X^{\mu}(z_{1})+
 		2\aph k_{1}\cdot\psi(z_{1})\psi^{\mu}(z_{1})\right) e^{ik_{1}\cdot X(z_{1})},
\end{equation}
and the vertex operator in the $(-1, -1)$ picture for the state at the level $n=1$ is
\begin{equation}
	W_{(-1,-1)}^{(1)}(k_2,z_{2},\bar{z_{2}})=G_{\rho\sigma\tau\zeta}V_{-1}^{\rho\sigma}(k_{2},z_{2})
		\widetilde{V}_{-1}^{\tau\zeta}(k_{2},\bar{z}_{2})
\end{equation} 
with 
\begin{equation}
	V_{-1}^{\rho\sigma}(k_{2},z_{2})=\dfrac{i}{\sqrt{2\alpha'}}e^{-\varphi(z_{2})}\partial X^{\rho}(z_{2})
	\psi^{\sigma}(z_{2})e^{ik_{2}\cdot X(z_{2})}.
\label{eq:level_one_holo}
\end{equation} 

We are interested in a scattering process in the presence of D-branes which break Lorentz invariance; this 
changes the kinematics and the boundary conditions (Neumann on the direction parallel to the brane, 
Dirichlet on the directions orthogonal to it). To implement these new conditions, we use the doubling 
trick. As reviewed in \cite{garousi96} in the context of D-brane physics, this simiplifies the treatment by substituting antiholomorphic 
fields $\bar{X}(\bz)$, $\tilde{\varphi}(\bz)$ which are functions of an antiholomorphic variable, with 
holomorphic fields depending on $\bz$ treated as an independent holomorphic variable. This is equivalent 
to sending $X^{\mu}(\bar{z}) \to {D^{\mu}}_{\nu}X^{\nu}(\bar{z})$, $\tilde{\psi}^{\mu}(\bar{z})\to 
{D^{\mu}}_{\nu}\psi^{\nu}(\bar{z})$ and $\widetilde{\varphi}(\bar{z}) \to \varphi(\bar{z})$
for correlation functions evaluated on $\mathbb{H}_+$, where $D=(\eta_{p+1}, -1_{9-p})$.
Having employed the doubling trick we need only compute correlators between the holomorphic fields. We will use
\begin{align}\nonumber
	\langle X^\mu(z)X^\nu(w)\rangle &= -2\aph\eta^{\mu\nu} \log(z-w),\\ \nonumber
	\langle \psi^\mu(z) \psi^\nu(w)\rangle &= \dfrac{\eta^{\mu\nu}}{z-w},\\
	\langle \phi(z)\phi(w)\rangle &= -\log(z-w).
\end{align}

There are several kinematic factors which appear in the following calculation which depend on the momenta 
carried by the holomorphic and antiholomorphic fields; we would like to be able to express 
these in terms of the variables above. From the definition of ${D^{\m}}_{\n}$ it can be seen that
the momenta satisfy $(D\cdot k)^{\m}=k_{\Vert}^{\m}-k_{\bot}^{\m}$ and so, with the aid of the conservation
of momentum, one can deduce the following identities,
\begin{align}
	k_1\cdot k_2 		&=	\frac{n}{2\aph}+\frac{n'}{2\aph}-\frac{t}{8}, \\
	k_1\cdot D\cdot k_1	&= 	-\frac{s}{2}+\frac{n}{\aph},\\
	k_2\cdot D\cdot k_2	&= 	-\frac{s}{2}+\frac{n'}{\aph}, \\
	k_1\cdot D\cdot k_2	&=	\frac{s}{2}+\frac{t}{8}-\frac{n}{2\aph}-\frac{n'}{2\aph}.
\end{align}\label{kin_identities}

Inserting the vertex operators~(\ref{eq:graviton_vertex_amp}-\ref{eq:level_one_holo}) and the 
normalization~(\ref{eq:normalisation}) into the integral~(\ref{eq:formal_amp}) we obtain
\beq
\begin{aligned}
 	 A_{0,1}=&\dfrac{\kappa T_p}{2}\int_{\mathbb{H}_+}\:\dfrac{d^2z_1d^2z_2}{V_{CKG}}\corl{W^{(0)}_{(0,0)}
		(k_1,z_1,\bz_1)	W^{(1)}_{(-1,-1)}(k_2,z_2,\bz_2)}_{\mathbb{H}_+}\\
	=&\dfrac{\k T_p}{8\aph^{\,2}}\,\ep_{\m \l}D^{\l}\:_{\nu}G_{\rho\sigma\alpha\beta}D^{\alpha}\:_{\tau}
		D^{\beta}\:_{\zeta}\int_{\mathbb{H}_+}\:\dfrac{d^2z_1d^2z_2}{V_{CKG}}\\
 	 &\bigg\langle :\big( i \partial X^{\mu}(z_{1})+2\aph k_{1}\cdot\psi(z_{1})\psi^{\mu}
		(z_{1})\big) e^{ik_{1}\cdot X(z_{1})}:\\
	 &:\big(i \bar{\partial} X^{\nu}(\bar{z}_{1})+2\aph k_{1}\cdot D\cdot \p(\bar{z}_{1})\p^{\n}(\bar{z}_{1})\big)  
		e^{ik_{1}\cdot D\cdot X(\bar{z}_{1})}:\\ 
	 & :e^{-\varphi(z_{2})}\partial X^{\rho}(z_{2})\psi^{\sigma}(z_{2})e^{ik_{2}\cdot X(z_{2})}:
	 	:e^{-\varphi(\bar{z}_{2})}\bar{\partial }X^{\tau}(\bar{z}_{2})\psi^{\zeta}(\bar{z}_{2})e^{ik_{2}\cdot D\cdot X(\bar{z}_{2})}:\bigg\rangle.
\end{aligned}
\label{eq:graviton_amp}
\eeq
In evaluating the above correlator it can be shown that the leading term in the high energy Regge limit
is given by the contraction of the two operators quadratic in the fermionic fields. This term is
proportional to $2\aph\, k_{1}\cdot D\cdot k_{1}=\aph s + 2 $ and the overall $s$ 
factor assures that this term is dominant with respect to all the other contractions in the amplitude.

In order to see that this is the case it is important to note the following fact. If we define an 
$SL(2,\mathbb R)$ invariant variable 
\beq
	\o = \frac{(\ztwo)(\zfive)}{(\zthree)(\zfour)}
\label{eq:omega}
\eeq
and use this in writing the factor in the correlation function which results from the contraction of the $e^{i k\cdot X}$ operators
\begin{equation}
	\left\langle e^{ik_{1}\cdot X(z_{1})}e^{ik_{1}\cdot D\cdot X(\bar{z}_{1})}e^{ik_{2}\cdot X(z_{2})}
		e^{ik_{2}\cdot D\cdot X(\bar{z}_{2})}\right\rangle=\o^{-\aph \tfrac{t}{4}+1}(\o-1)^{-\aph s}
		(\zsix)^{2},
\end{equation}
then the remaining explicit $z$-dependence can be combined with that from the other possible contractions 
in (\ref{eq:graviton_amp}), together with the appropriate measure $\frac{d^2z_1d^2z_2}{V_{CKG}}\mapsto d\o
(\zthree)^2(\zfour)^2$, to give some multiplicative $SL(2,\mathbb R)$ invariant function $F(\o)$. 
The amplitude then takes the following schematic form
\beq
	A_{0,1}=\dfrac{\k T_p}{2}\int_0^1d\o\, \o^{-\aph \tfrac{t}{4}+1}(\o-1)^{-\aph s}F(\o).
\label{eq:invariant_amp}
\eeq
The behaviour of this integral when we take the limit $\aph s \to \infty$ is controlled by the integrand
in the neighbourhood of the point $\o=0$ and since $F(\o)$ is, in general, a sum of terms composed of 
powers of $\o$, $(1-\o)$ and their inverse quantities, $A_{0,1}$ itself consists of a sum of integrals
of the form shown below,
\beq
	\int_0^1d\o\, \o^{-\aph \tfrac{t}{4}+a}(1-\o)^{-\aph s +b}=\frac{\G \left(-\aph \tfrac{t}{4}+a+1\right ) 
		\G \left(-\aph s+b+1\right )}{\G \left(-\aph \tfrac{t}{4}-\aph s+a+b+2\right )}.
\eeq
It can be seen that in the high energy limit this quantity will scale with $s$ as $(\aph s)^{-a-1}$ and 
therefore the dominant contribution to $A_{0,1}$ will come from the term with the lowest value of $a$.
The definition of $\o$ in equation~(\ref{eq:omega}) implies that these both point to large $s$ 
behaviour being governed by the region of integration over the world sheet in which the two vertex 
operators are brought together, that is to say when $z_1\to z_2$ and $\bz_1\to \bz_2$. The analysis of 
this process will be expanded upon and made more systematic in the next Section.

Having learnt this, one can quickly deduce that we are interested in the terms of $F(\o)$ which are 
obtained from the maximum possible number of contractions between the holomorphic and antiholomorphic
fields in equation~(\ref{eq:graviton_amp}). In general there are many such terms but, as mentioned below
(\ref{eq:graviton_amp}), those which contain the contraction $\contraction[0.5ex]{k_1\cdot}{\psi}{\:k_1
\cdot D \cdot}{\psi}\nomathglue{k_1\cdot \psi \:k_1\cdot D\cdot\psi}$ 
will bring an additional factor $\aph s$ and so will ultimately be the leading terms at high energy.

By evaluating the correlation function in equation~(\ref{eq:graviton_amp}), then employing the physical 
state conditions $\ep_{\m \n}k_1^{\m}=G_{\rho\sigma\tau\zeta} k_{2}^{\r}=0$ and momentum conservation
$k_{1}^{\m}+(D\cdot k_{1})^{\m}+ k_{2}^{\m}+(D\cdot k_{2})^{\m}=0$,
we can determine the function $F(\o)$ and perform the integral in equation~(\ref{eq:invariant_amp}). 
If we take the limit $\aph s\rightarrow\infty$ we find that only one term dominates in this case, which is 
proportional to $k_1^\r G_{\r \s \tau\zeta}\:\ep^{\s \tau} k_1^\zeta$. 
Subsequent application of Stirling's approximation for the Gamma function yields the following form for
the amplitude in the high energy limit
\begin{equation}
	A_{0,1} \sim  \frac{\kappa T_p}{2} e^{-i\pi \alpha' \tfrac{t}{4}}\Gamma\left (-\tfrac{\alpha' t}{4}\right )
		\left ( \alpha' s\right)^{\tfrac{\alpha' t}{4}+1}\frac{\alpha'}{2} q^\rho  G_{\rho\sigma\tau\zeta}
		\:\epsilon^{\sigma\tau} q^\zeta,
\label{leading_ampl1}
\end{equation} 
where we have replaced $k_1$ and $k_2$ with the physical momentum transferred $q=2(k_1+k_2)$ by virtue of
the physical state conditions. 

It is instructive to compare this result with the analogous result for the elastic scattering of a 
graviton by a D-brane,
\beq
	A_{0,0} \sim  \frac{\kappa T_p}{2} e^{-i\pi \aph \tfrac{t}{4}}\Gamma\left (-\tfrac{\aph t}{4}\right )
		\left ( \aph s\right)^{\tfrac{\aph t}{4}+1} {\ep_1}_{\s \tau}
		\:\ep_2^{\s \tau}. 
\eeq
Considered purely as functions of the complex variables $s$ and $t$, these two equations demonstrate 
identical behaviour; the physical amplitude is obtained by taking $s$, $t$ real with large positive $s$
and negative $t$, but for positive $t$ we see that there are poles corresponding to the exchange of a
string of mass $M^2=t$ with the D-brane and, futhermore, the $s$-dependence 
indicates that the exchanged string has spin $J=(\aph M^2+4)/2$ --- it belongs to the Regge trajectory of 
the graviton. As the energy becomes very large this description of the exchange of a single string breaks
down due to the violation of unitarity, however, as demonstrated in \citep{d'appollonio10} unitarity may
be recovered by taking into account loop effects. Examining the two amplitudes above we note that
 the elastic and the inelastic amplitude
only differ in the multiplicative factor containing the initial and final polarisation tensors and
the exchanged momentum.

In the next Section we shall generalise this result and compute the leading
high energy contribution to scattering amplitudes involving two generic states of the leading Regge 
trajectory, at mass levels $n$ and $n'$ of the spectrum; to do this we shall make use of the OPE for the
two vertex operators to isolate those terms in the amplitude which dominate as the vertices are brought
together on the world sheet.
%
\section{Computation via OPE methods}
\label{sec:ope_methods}
%
It is seen in section \ref{sec:amp_comp} that the usual methods used for calculating scattering 
amplitudes in string theory can result in the generation of a vast number of terms which transpire
to be subleading in energy after integration over the world-sheet. If our interests lie only in the
leading terms then we would like to be able to distinguish these subleading terms and discard them at
the very beginning of our calculation; with OPE methods we can do this quite easily and, furthermore,
they highlight the simple structure present in the class of amplitudes we consider here.

To compute the leading terms in the amplitude given by equation~(\ref{eq:formal_amp}) in the Regge limit, 
$\aph s \to \infty$ and $\aph t$ fixed, it is key to note that the integral over the world-sheet in 
(\ref{eq:formal_amp}) is dominated at large $\aph s$ by the behaviour of the vertex operators as $z_1 \to
z_2$, as exemplified in equation~(\ref{eq:invariant_amp}). Specifically, when working in the Regge regime 
for scattering processes with two external string states the leading contribution to the integral over the 
world-sheet can be taken from the OPE of the vertex operators. Subsequent integration over their separation 
$w=z_1-z_2$ will give an amplitude with the expected Regge behaviour; however, care must be taken in this 
process, as we will see, since there exist terms subleading in $w$ which will contribute factors of 
$\aph s$ and by doing so prevent us from neglecting them.

\begin{figure}
	\begin{center}
		\includegraphics[scale=0.7]{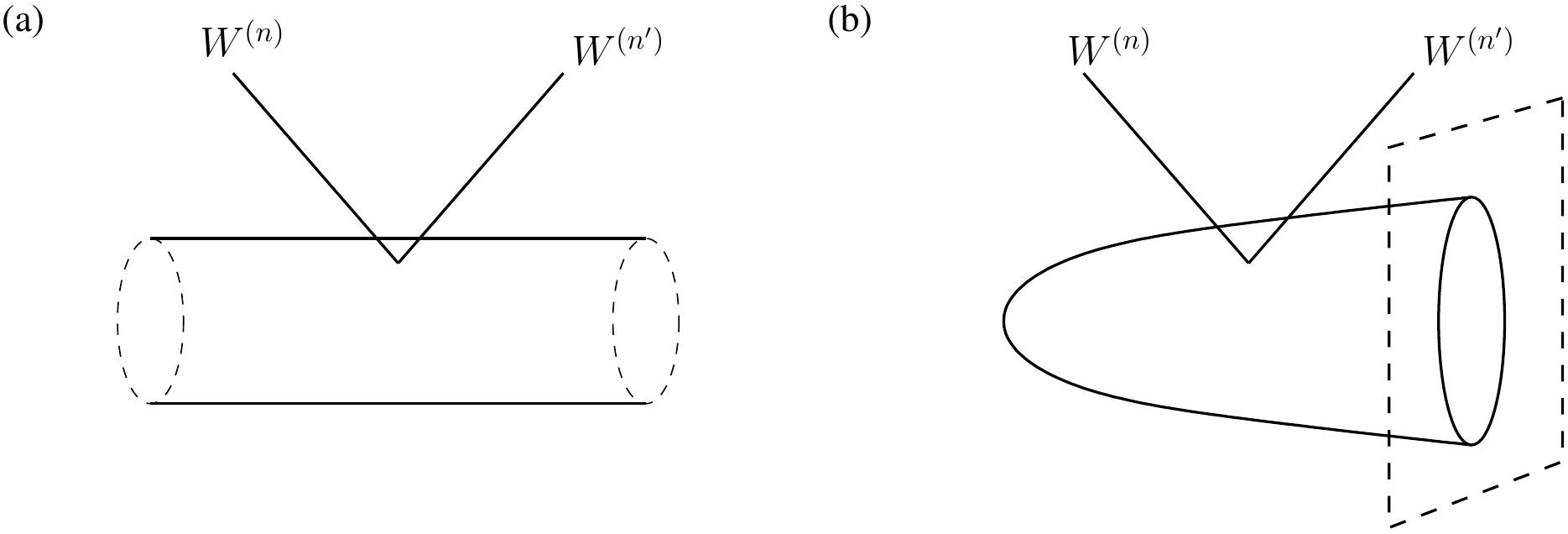}
	\end{center}
	\caption{(a) The pomeron vertex operator given by two physical vertices $W^{n}$ and $W^{n'}$ on the infinite cylinder. (b)
		 The addition of a boundary gives the process in which these states interact with a D-brane.}
	\label{fig:boundary_state}
\end{figure}

Rather than evaluating the correlator in (\ref{eq:formal_amp}), integrating the result and 
then computing the asymptotic form for large $\aph s$ we may instead determine the OPE of 
the two vertex operators and integrate out the dependence on the separation $w=z_1-z_2$. 
This process yields a quasi-local operator referred to as the pomeron vertex operator, first introduced in 
\citep{ademollo89} as the reggeon and recently used in \citep{brower06, fotopoulos10}. 
As illustrated in figure~\ref{fig:boundary_state}(a), the process of constructing this operator involves 
taking the limit in which two physical vertices approach
one another on a surface topologically equivalent to the infinite cylinder, hence it can be considered to 
properly describe the $t$-channel exchange of a closed string. The pomeron vertex operator itself, 
up to subleading term in $s$, also 
satisfies the physical state conditions for any $\aph t$, and for $\aph t=4n$ ($n=0,1,\ldots$) we can 
think of it as describing the exchange of a string with mass given by $\aph M^2=4n$. As explained in 
\citep{brower06} the importance of single pomeron exchange lies in the fact that it dominates scattering 
amplitudes in the Regge regime both in QCD for $N_c\rightarrow\infty$ and in string theory.

The information contained within the pomeron vertex is only that of the two physical states which 
produce the exchanged pomeron and the restriction of our kinematic invariants to the Regge limit, as such
this vertex may be used to generate many amplitudes of interest by insertion onto the appropriate 
world-sheet. In this instance we are interested in a world-sheet with a single boundary and boundary
conditions which correspond to the presence of a D$p$-brane, indicated in figure~\ref{fig:boundary_state}(b). 
This can be easily done using boundary state methods which were introduced in \citep{callan86} and are 
reviewed in \citep{divecchia99}. These methods are extremely powerful for dealing with problems involving 
string interactions with D-branes \citep{duo07,black10} and for our purposes we can emulate the effects of
the boundary state using the doubling trick, but further applications using the same vertex can be carried out 
systematically by bearing these facts in mind.

We will illustrate the use of the pomeron vertex operator with some specific examples before proceeding 
to the general case. If we label by 
$(n,n')$ the amplitude in which a state of mass $\aph M_1^2=4n$ undergoes a transition to a state of mass 
$\aph M_2^2=4n'$ then the examples we shall consider will be $(0,n')$ and $(1,n')$. At the end of this section
we will generalize our results to the case $(n, n')$. 
%
\subsection{Graviton-massive state transitions}
%
In this case we examine the processes in which a massless state is excited to another of arbitrary mass.
 The physical polarizations for massive states will be written as tensor products of their holomorphic and antiholomorphic 
components,
\begin{align}
\epsilon_{\mu_1\ldots\mu_{n} \alpha\:\nu_1\ldots \nu_{n} \beta}&= \varepsilon_{\mu_1\ldots\mu_{n} \alpha} \otimes\widetilde{\varepsilon}_{\nu_1\ldots \nu_{n} \beta} ,
\\
G_{\mu_1\ldots\mu_{n} \alpha\:\nu_1\ldots \nu_{n} \beta}&= \mathcal{G}_{\mu_1\ldots\mu_{n} \alpha} \otimes\widetilde{\mathcal{G}}_{\nu_1\ldots \nu_{n} \beta} .
\end{align}

Our first task will be to determine the OPE of the vertex operators $W_{(0,0)}^{(0)}(k_1,z_1,\bz_1)$ and 
$W_{(-1,-1)}^{(n)}(k_2,z_2,\bz_2)$ as $z_1\to z_2$ and $\bz_1\to\bz_2$; this task may seem daunting at 
first due to the large number of possible contractions, but as we shall see it is possible to immediately 
identify which contractions will end up giving the leading order contribution to this amplitude. In this 
particular computation our vertex operators are
\beq
\begin{aligned}
	W_{(0,0)}^{(0)}(k_1,z_1,\bz_1)&=-\epsilon_{\m \n}V^{\m}_0(k_1,z_1)\widetilde{V}_0^{\n}(k_1,\bz_1), \\
	V_0^{\m}(k_1,z_1)&=\frac{1}{\sqrt{2\aph}}\left(i\part X^{\m}(z_1)+2\aph k_1\cdot \p(z_1) \p^{\m}(z_1)\right)e^{ik_1\cdot X(z_1)},
\end{aligned}
\label{eq:graviton_vertex}
\eeq
and
\beq
\begin{aligned}
	W_{(-1,-1)}^{(n')}(k_2,z_2,\bz_2)&=\mathcal{G}_{\r_1\ldots\r_{n'} \s}\widetilde{\mathcal{G}}_{\l_1\ldots \l_{n'} \g}
		V^{\r_1\ldots\r_{n'}\s}_{-1}(k_2,z_2)\widetilde{V}_{-1}^{\l_1\ldots \l_{n'}\g}(k_2,\bz_2), \\
	V^{\r_1\ldots\r_{n'}\s}_{-1}(k_2,z_2)&=\frac{1}{\sqrt{n'!}}\left(\frac{i}{\sqrt{2\aph}}\right)^{n'}e^{-\varphi(z_2)}\prod_{i=1}^{n'}\part X^{\r_i}\p^{\s}
	e^{ik_2\cdot X(z_2)}.
\end{aligned}
\label{eq:graviton_nvertex}
\eeq
In deriving the pomeron vertex operator we consider first the insertion of the above two closed string 
vertices onto a world-sheet with the topology of the Riemann sphere. We need not consider the contractions
between holomorphic and antiholomorphic operators until the resulting effective vertex is inserted onto a 
world-sheet with the topology of the disc.
As such we can write the resulting OPE in terms of the world-sheet separation $w=z_1-z_2$ and the point 
$z=\tfrac{z_1+z_2}{2}$ which then takes the following form
\beq
	W_{(0,0)}^{(0)}\Big(k_1,z+\tfrac{w}{2},\bz+\tfrac{\bw}{2}\Big)W_{(-1,-1)}^{(n')}\Big(k_2,z-\tfrac{w}{2},\bz-\tfrac{\bw}{2}\Big)\sim 
		-|w|^{-\aph \tfrac{t}{2}+2n'}\mathcal{O}(z,w)	\widetilde{\mathcal{O}}(\bz,\bw).
	\label{eq:OPE}
\eeq
It is simple to check that the operators $\mathcal{O}$ and $\widetilde{\mathcal{O}}$ are polynomials 
of at most degree $(n'+1)$ in $w^{-1}$ and $\bw^{-1}$ respectively, with an exponential factor contributing 
terms subleading in the small $w$ limit, that is
\beq
	\mathcal{O}(z,w)=e^{i\tfrac{1}{2}(k_1-k_2)\cdot \part X(z)w}\sum_{p=1}^{n'+1}\frac{\mathcal{O}_p(z)}{w^p}, \quad 
		\widetilde{\mathcal{O}}(\bz,\bw)=e^{i\tfrac{1}{2}(k_1-k_2)\cdot \bar{\part} X(\bz)\bw}\sum_{q=1}^{n'+1}
		\frac{\widetilde{\mathcal{O}}_q(\bz)}{\bw^q}.
	\label{eq:w_expansion}
\eeq
In the high energy limit it is necessary to retain these particular subleading terms in the exponential, as 
we will see, because contractions between $k_1\cdot \part X$ and $k_1 \cdot \bar{\part}X$ will generate 
factors of $s$ meaning that these terms cannot be neglected for $|w|^2\sim(\aph s)^{-1}$. It is in fact 
these terms which will generate the Regge behaviour that we expect.

The momentum exchanged between the string and the brane may be written as $q=p_1+p_2=2(k_1+k_2)$, hence
$-q^2=t$, and it is also useful to define a vector $\tilde{q}=2(k_1-k_2)$.
In the expansions given by equation~(\ref{eq:w_expansion}) it will be the most singular terms which will 
dominate in the pomeron vertex operator, this operator being obtained by the integration of 
$w$ in the OPE (\ref{eq:OPE}) over the complex plane, and this procedure will in general result in an 
integral for each of these terms of the form
\beq \label{integral}
	\int_{\mathbb{C}}d^2w\, |w|^{-\aph \tfrac{t}{2}-2}e^{i\tfrac{\tilde{q}}{4}\cdot \part X(z)w}e^{i\tfrac{\tilde{q}}{4}
	\cdot \bar{\part}X(\bz)\bw}.
\eeq
The integration of (\ref{integral}) can be done by introducing new variables
$u=\frac{\tilde{q}}{4}\cdot \part X(z)\:w$, $\bar{u}=\frac{\tilde{q}}{4}\cdot \bar{\part}X(\bz)\:\bw$,
\beq
	e^{-i\pi\aph\frac{t}{4}}\int_{\mathbb{C}}d^2u\, |u|^{-\aph \tfrac{t}{2}-2}\:e^{i(u+\bar{u})}\left(i\frac{\tilde{q}}{4}\cdot \part X(z)
		\right)^{\aph \tfrac{t}{4}}\left(i\frac{\tilde{q}}{4}\cdot \bar{\part}X(\bz)\right)^{\aph \tfrac{t}{4}}.
\eeq
Then we integrate over the positions $u=r e^{i\theta}$ using the following integrals:
\begin{align}
	&\int_0^{2\pi}d\theta\:e^{-2ir\cos\theta}=2\pi\:J_0(2r), \\
	\label{Bessel}
	&\int_0^\infty\:dr\:r^a\:J_0(2r)=\frac{1}{2}\frac{\Gamma\left(\frac{1+a}{2}\right)}{\Gamma\left(\frac{1-a}{2}\right)},
\end{align}
and we get
\beq
\int_{\mathbb{C}}d^2w\, |w|^{-\aph \tfrac{t}{2}-2}e^{i\tfrac{\tilde{q}}{4}\cdot \part X(z)w}e^{i\tfrac{\tilde{q}}{4}
	\cdot \bar{\part}X(\bz)\bw}
	=\Pi(t)
         \left(i\frac{\tilde{q}}{4}\cdot \part X(z)\right)^{\aph \tfrac{t}{4}}\left(i\frac{\tilde{q}}{4}\cdot \bar{\part}X(\bz)\right)^{
	\aph \tfrac{t}{4}},
\eeq
where $\Pi(t)$ is commonly referred to as the pomeron propagator \citep{brower06,fotopoulos10} and is given 
by the following
\beq \label{pomeron_prop}
	\Pi(t)=	2\pi\frac{\G\left(-\aph \tfrac{t}{4}\right)}{\G\left(1+\aph\tfrac{t}{4}\right)}\:e^{-i \pi \aph \tfrac{t}{4}}.
\eeq
The two-point function is reduced to a one-point function  of the effective pomeron vertex on the disc.

From what we have discussed so far, one can conclude that to leading order in energy the pomeron vertex 
operator should take the following form
\begin{equation} \label{integral_2}
	\int_{\mathbb{C}}d^2w\, W_{(0,0)}^{(0)}(k_1,z+\tfrac{w}{2},\bz+\tfrac{\bw}{2})W_{(-1,-1)}^{(n')}(k_2,z-\tfrac{w}{2},\bz-\tfrac{\bw}{2})
\sim
	-K_{0,n'}(q,\ep,G) \Pi(t)\mathcal{O}(z)\widetilde{\mathcal{O}}(\bz),
\end{equation}
where we have the pomeron propagator, the normal-ordered operators
\begin{subequations}\label{O_Otilde}
\begin{align}
	\mathcal{O}(z)&=\sqrt{2\aph}\left(i\frac{\tilde{q}}{4}\cdot \part X\right)^{\aph \tfrac{t}{4}}k_1\cdot \p e^{-\varphi}e^{i \tfrac{q}{2}\cdot X}, \\
	\widetilde{\mathcal{O}}(\bz)&=\sqrt{2\aph}\left(i\frac{\tilde{q}}{4}\cdot \bar{\part} X\right)^{\aph \tfrac{t}{4}}k_1\cdot \pt e^{
	-\widetilde{\varphi}}e^{i\tfrac{q}{2}\cdot X},
\end{align}
\end{subequations}
and a kinematic function, dependent upon the polarisation tensors and the momentum transferred from 
the string to the brane, 
\beq
	K_{0,n'}(q,\ep,G)=\frac{1}{n'!}\left(\frac{\aph}{2}\right)^{n'}\left(k_1\right)^{n'}\cdot \mathcal{G}_0\cdot \varep_0\cdot \left(k_2\right)^{n}\, 
	\left(k_1\right)^{n'}\cdot \widetilde{\mathcal{G}}_0\cdot \widetilde{\varep}_0\cdot \left(k_2\right)^0.
\eeq
Here we have introduced for later convenience the notation  
\beq
	\left(k_1\right)^{n'}\cdot\mathcal{G}_a\cdot \varep_a\cdot \left(k_2\right)^n =2^{n+n'-a-b}\left (\prod_{i=1}^{a}\eta^{\r_i \m_i} \right )\left 
		(\prod_{j=a+1}^{n'}k_1^{\r_j}\right ) \left (\prod_{k=a+1}^{n} k_2^{\m_k}\right ) \mathcal{G}_{\r_1\ldots \r_{n'}\s},
		\eta^{\s \ah}\varep_{\m_1\ldots \m_n\ah}
	\label{eq:dot_product}
\eeq
where  $a\in\{0,\ldots,\min \{n,n'\}\}$ will count the number of contractions between 
$\mathcal{G}$ and $\varep$ in addition to that arising from the fermionic fields, and an analogous 
expression holds for the polarisation of the antiholomorphic 
components. In (\ref{eq:dot_product}) products of the form $\prod_{i=1}^0$ should be replaced by unity.
 This notation will prove useful since the polarisation tensors for all states on the leading 
Regge trajectory are symmetric in all holomorphic indices and symmetric in all antiholomorphic indices, 
as a result the order of contractions with these indices is immaterial; all we need do is keep track 
of how many factors of $k_1$ are contracted with $\mathcal{G}$, $\widetilde{\mathcal{G}}$ and how many 
factors of $k_2$ are contracted with $\varep$, $\widetilde{\varep}$. Furthermore, due to the requirement 
that longitudinal polarisations vanish we find that we may replace all occurrences of $k_1$ and $k_2$ in 
equation~(\ref{eq:dot_product}) and its antiholomorphic partner with the transferred momentum $q=2(k_1+k_2)$.
In this case we can substitute the expression in equation~(\ref{eq:dot_product}) with
\beq
	q^{n'}\cdot\mathcal{G}_a\cdot \varep_a\cdot q^n =\left (\prod_{i=1}^{a}\eta^{\r_i \m_i} \right )\left (\prod_{j=a+1}^{n'}q^{\r_j}\right ) 
		\left (\prod_{k=a+1}^{n} q^{\m_k}\right ) \mathcal{G}_{\r_1\ldots \r_{n'}\s}
		\eta^{\s \ah}\varep_{\m_1\ldots \m_n\ah}.
\eeq
The one-point function of the pomeron vertex is given by the contraction of the operators 
$\mathcal{O}(z)$, $\widetilde{\mathcal{O}}(\bz)$ in (\ref{O_Otilde}). To leading order in $s$, discarding 
the masses, we obtain
\begin{multline}
	K_{0,n'}(q,\ep,G) \int_{D_2} \dfrac{d zd\bz}{V_{CKG}}\Pi(t)\langle:\mathcal{O}(z)::\widetilde{\mathcal{O}}(\bz):\rangle_{D_2}
	 \sim K_{0,n'}(q,\ep,G)(\aph s)^{\aph\frac{ t}{4}+1}\\
	\times \frac{\Pi(t)}{2\pi}\Gamma\Big(1+\aph\tfrac{t}{4}\Big).
\end{multline}
The factor $\frac{1}{2\pi}$ arises from the ratio between the integration over the insertion point of the vertex operator and
the volume of the Conformal Killing Group of the disc,  $SL(2,\mathbb{R})$.

We obtain an \sl invariant function which gives the
leading high energy behaviour of the amplitude for a graviton scattering from a D-brane into a
state on the leading Regge trajectory at level $n'$,
\beq
	A_{0,n'}(s,t)=\frac{\k T_p}{2}K_{0,n'}(q,\ep,G) \G(-\aph \tfrac{t}{4})e^{-i\pi \aph \tfrac{t}{4}} \,
	(\aph s)^{\aph \tfrac{t}{4}+1}.
	\label{eq:zeroton_amp}
\eeq
%
\subsection{Transitions from the lowest massive state}
\label{sec:1ton}
%
We next move on to the case of a string with mass $M_1^2=4/\aph$ interacting with a D-brane to leave a 
string of mass $M_2^2=4n'/\aph$. The vertex operators are now given by
\beq
\begin{aligned}
	W_{(0,0)}^{(1)}(k_1,z_1,\bz_1)&=-\varep_{\m \ah}\widetilde{\varep}_{\n \b}V^{\m \ah}_0(k_1,z_1)
		\widetilde{V}_0^{\n \b}(k_1,\bz_1), \\
	V_0^{\m \ah}(k_1,z_1)&=\frac{i}{2\aph}\left(i\part X^{\m}\part X^{\ah}+2\aph k_1\cdot \p \p^{\ah}\part X^{\m}
	-i2\aph \part \p^{\m} \p^{\ah}\right)e^{ik_1\cdot X},
\end{aligned}
\eeq
and
\beq
\begin{aligned}
	W_{(-1,-1)}^{(n')}(k_2,z_2,\bz_2)&=\mathcal{G}_{\r_1\ldots\r_{n'} \s}\widetilde{\mathcal{G}}_{\l_1\ldots \l_{n'} \g}
		V^{\r_1\ldots\r_{n'}\s}_{-1}(k_2,z_2)\widetilde{V}_{-1}^{\l_1\ldots \l_{n'}\g}(k_2,\bz_2), \\
	V^{\r_1\ldots\r_{n_2}\s}_{-1}(k_2,z_2)&=\frac{1}{\sqrt{n'!}}\left(\frac{i}{\sqrt{2\aph}}\right)^{n'}e^{-\varphi(z_2)}\prod_{i=1}^{n'}\part X^{\r_i}\p^{\s}
	e^{ik_2\cdot X(z_2)}.
\end{aligned}
\eeq

The methods developed in the previous example need little modification in order to deal with this problem. 
With them we can easily determine the form of the pomeron vertex operator and it can be written in the 
same manner as in the $(0,n')$ case,
\beq
	\int_{\mathbb{C}}d^2w\, W_{(0,0)}^{(1)}(k_1,z+\tfrac{w}{2},\bz+\tfrac{\bw}{2})W_{(-1,-1)}^{(n')}(k_2,z-\tfrac{w}{2},\bz-\tfrac{\bw}{2})\sim 
	-K_{1,n'}(q,\ep,G) \Pi(t)\mathcal{O}(z)\widetilde{\mathcal{O}}(\bz),
	\label{eq:oneton_pomeron}
\eeq
where this time
\begin{multline}
  K_{1,n'}(q,\ep,G)=\frac{1}{n'!} \left(\frac{\aph}{2}\right)^{n'+1}\bigg [q^{n'}\cdot\mathcal{G}_0\cdot \varep_0\cdot q^1 \,q^{1}\cdot\widetilde{\varep}_0 \cdot 
  \widetilde{\mathcal{G}}_0\cdot q^{n'}
  -\frac{2n'}{\aph}q^{n'-1}\cdot\mathcal{G}_{1}\cdot \varep_1\cdot q^0 \, q^1\cdot\widetilde{\varep}_0 \cdot \widetilde{\mathcal{G}}_0\cdot q^{n'} \\
  -\frac{2n'}{\aph}q^{n'}\cdot\mathcal{G}_{0}\cdot \varep_0\cdot q^1 \, q^0\cdot \widetilde{\varep}_1 \cdot \widetilde{\mathcal{G}}_{1}\cdot q^{n'-1} 
  +\left(\frac{n'}{2\aph}\right )^2q^{n'-1}\cdot \mathcal{G}_{1}\cdot \varep_1 \cdot q^0 \, q^0\cdot \widetilde{\varep}_1 \cdot \widetilde{\mathcal{G}}_{1} \cdot q^{n'-1} \bigg ]
	\label{eq:kinematic_function}
\end{multline}
and all other quantities remain as previously defined. Because of this, the resulting amplitude will be 
identical to that of equation~(\ref{eq:zeroton_amp}) other than the form of the kinematic function which 
is given in (\ref{eq:kinematic_function}), 
\beq
	A_{1,n'}(s,t)=\frac{\k T_p}{2}K_{1,n'}(q,\ep,G)\:\G(-\aph \tfrac{t}{4})
	e^{-i\pi \aph \tfrac{t}{4}} \,	(\aph s)^{\aph \tfrac{t}{4}+1}.
\eeq
Note that we have taken $2\aph k_1\cdot D\cdot k_1\sim -\aph s$ in the large $s$ limit, neglecting a mass 
term of the order of unity. In the next case we will move on to masses which contribute terms of order
$n$, $n'$ and we reiterate that we shall be considering only those states for which the rest mass 
contribution to the total energy is negligible. 
%
\subsection{Transitions within the leading Regge trajectory}
%
Here we return to our original consideration, the process in which a string in a state on the leading 
Regge trajectory is scattered from a D$p$-brane into some other state on the leading Regge trajectory. The 
vertex operators are those given by equations (\ref{eq:vertex_1}) and (\ref{eq:vertex_2}), with 
polarisation tensors $\varep \otimes \widetilde{\varep}$ and $\mathcal{G}\otimes \widetilde{\mathcal{G}}$,
respectively, and our methods 
will imitate those we have seen already. If we initially suppose $n$ and $n'$ to be fixed at some finite 
values, then by identifying the terms in the OPE of these vertices which will 
give the leading high energy contributions and integrating out the dependence on their separation we 
obtain the same general form for the pomeron vertex operator as in section~\ref{sec:1ton},
\beq
	\frac{1}{n!n'!}	\int_{\mathbb{C}}d^2w\, W_{(0,0)}^{(n)}(k_1,z+\tfrac{w}{2},\bz+\tfrac{\bw}{2})
	W_{(-1,-1)}^{(n')}(k_2,z-\tfrac{w}{2},\bz-\tfrac{\bw}{2})\sim -K_{n,n'}(q,\ep,G) \Pi(t)\mathcal{O}(z) \widetilde{\mathcal{O}}(\bz),
\eeq
where now we can see the full structure of the kinematic function, which may be written as the product of 
a contribution from the holomorphic operators with a contribution from the antiholomorphic operators, 
\begin{multline}
	K_{n,n'}(q,\ep,G)=\frac{1}{n!n'!}\left(\frac{\aph}{2}\right)^{n+n'}\sum_{a,b=0}^{\min \{n,n'\}}\left(-\frac{\aph}{2}\right)^{-a-b}C_{n,n'}(a) C_{n,n'}(b) \\
\times q^{n'-a}\cdot\mathcal{G}_a\cdot \varep_a\cdot q^{n-a}\,q^{n-b}\cdot \widetilde{\varep}_b \cdot \widetilde{\mathcal{G}}_b\cdot q^{n'-b}.
	\label{eq:polarisation}
\end{multline}
Here $\min\{n,n'\}$ indicates the smallest value from the set $\{n,n'\}$, and the combinatorial factors 
$C_{n,n'}$ (one each coming from the holomorphic and antiholomorphic contractions) come from the large 
number of possible contractions in the OPE which lead to the same operator after taking into account the 
symmetry of the polarisation tensors. These functions are given by
\beq
	C_{n,n'}(p)=\frac{n! n'!}{p! (n-p)! (n'-p)!}
\eeq
and this can be deduced in the following manner. If we consider the contribution from the holomorphic
operators then $K_{n,n'}$ is determined by all possible contractions amongst 
\beq
	\varepsilon_{\mu_1\ldots\mu_{n} \alpha}\mathcal{G}_{\rho_1\ldots\rho_{n'} \sigma}
	:\prod_{i=1}^{n}\partial X^{\m_i}e^{ik_1\cdot X}::\prod_{i=1}^{n'}\partial X^{\r_i}e^{ik_2\cdot X}:.
\eeq
Since $\varep$ and $\mathcal{G}$ are symmetric in all indices these contractions can simply be labelled by
the number of contractions beween $\partial X^{\m_i}$ and $\partial X^{\r_i}$, let this be $a$. This being
the case we must count how many ways one can generate $a$ such contractions, first one must choose $a$ 
operators from a total of $n$ possibilities for which there are $\tbinom{n}{a}=n!/a!(n-a)!$ different 
choices. Similarly we must choose a further $a$ operators from a set of $n'$ possibilities giving 
another factor of $\tbinom{n'}{a}$. Finally, from this set of $2a$ operators there $a!$ possible ways to 
contract them in pairs. The product of these numbers gives the total number of different contractions 
which result in a factor $\mathcal{G}_a\cdot \varep_a$ and is equal to the function $C_{n,n'}(a)$.

With this new form for the kinematic factor $K$ we are finished, the rest of the computation having 
already been solved in the previous two examples. The final result for the amplitude of a finite mass 
string scattering from a D-brane is 
\beq
	A_{n,n'}(s,t)=\frac{\k T_p}{2} K_{n,n'}(q,\ep,G)\: \G\left(-\aph \tfrac{t}{4}\right)
	e^{-i\pi \aph \tfrac{t}{4}}\,(\aph s)^{\aph \tfrac{t}{4}+1}.
	\label{eq:result_mn}
\eeq
%
\section{Comparison with the eikonal analysis}
\label{sec:eikonal}
%
In a recent work \cite{d'appollonio10} it has been shown that tree-level amplitudes of closed strings in 
the background of $N$ D$p$-branes exponentiate to give the S-matrix an operator eikonal form at high 
energy, 
\beq
	S(s,\mathbf{b})=e^{i2\hat{\d}(s,\mathbf{b})}, \quad \quad 2\hat{\d}(s,\mathbf{b})=\frac{1}{2\sqrt{s}}\int_0^{2\pi}\frac{d\s}{2\pi}
	:\mathcal{A}\left (s,\mathbf{b}+\hat{\mathbf{X}}(\s)\right ):,
	\label{eq:eikonal}
\eeq
thus generalising the field theory result for the S-matrix. Here, $\mathcal{A}(s,\mathbf{b})$ can be 
obtained from the Regge limit of the disk amplitude for the elastic scattering of a graviton from the 
D$p$-branes after stripping it of its dependence upon the polarisation,
\beq
	A(s,\mathbf{q})=\frac{\k T_p}{2}\G\left(\aph \frac{q^2}{4}\right)
	e^{i\pi \aph \tfrac{q^2}{4}}\,(\aph s)^{-\aph \tfrac{q^2}{4}+1},
\label{eq:stripped_amp}
\eeq
by performing a Fourier transformation from the space of transverse momenta $q$ to that of impact 
parameter $b$,
\beq
	\mathcal{A}(s,\mathbf{b})=\int \frac{d^{8-p}q}{(2\pi)^{8-p}}\, A(s,\mathbf{q})e^{i \mathbf{b}\cdot\mathbf{q}}.
\eeq
In shifting the impact parameter by the string position operator $\hat{X}$, as in \citep{amati87a}, one can 
take into account the finite size of the string and it is in this respect that the eikonal operator of 
string theory differs from the field theory eikonal. It should be possible to derive the amplitudes 
evaluated by direct computation in the previous sections by using the eikonal operator.
One should note that in the following analysis neither $A(s,\mathbf{q})$
nor its Fourier transform $\mathcal{A}(s,\mathbf{b})$ need be specified and so these arguments are not 
restricted to any particular kinematic regime beyond that already assumed for the Regge limit.

The eikonal operator depends only on the
bosonic modes associated to the directions transverse to the brane and therefore its
action on the bosonic modes associated to the directions parallel to the brane
and on the fermionic modes is trivial. This is true also for the high energy
limit of the tree-level amplitudes, as emphasized in the previous sections.
In this section we shall show that the matrix elements of the eikonal operator
coincides precisely with the high energy limit of the tree-level string amplitudes. Once we are satisfied 
that this is the case we will consider the impact parameter space representation of the 
result~(\ref{eq:result_mn}) in the limit of large impact parameter, equivalent to small momentum transfer.

The exponential in (\ref{eq:eikonal}) represents a resummation of the perturbative expansion and we expect 
that tree level diagrams arise from the linear term in this exponential; that is to say we should be able 
to obtain amplitudes for the high energy limit of tree level processes from matrix elements of the operator 
$\bar{\mathcal{A}}$ defined to be
\beq
	\bar{\mathcal{A}}\equiv \int_0^{2\pi}\frac{d\s}{2\pi}:\mathcal{A}\left(s,\mathbf{b}+\mathbf{X}(\s)\right): =\sum_{k=0}^\infty
	\frac{1}{k!}\frac{\part^k \mathcal{A}(s,\mathbf{b})}{\part b^{\m_1}\ldots \part b^{\m_k}}\overline{X^{\m_1}\ldots X^{\m_k}}.
\eeq
To demonstrate this technique we will first reproduce the tree-level graviton to graviton scattering 
amplitude, before moving on to the more general case. The initial and final states representing the graviton are:
\begin{subequations}
\label{gravitons}
\begin{align}
 	\ket{i}&=\epsilon_{1\mu\nu}\psi^\mu_{-\frac{1}{2}}\widetilde{\psi}^\nu_{-\frac{1}{2}}\ket{0;0}, \\
 	\ket{f}&=\epsilon_{2\rho\sigma}\psi^\rho_{-\frac{1}{2}}\widetilde{\psi}^\sigma_{-\frac{1}{2}}\ket{0;0}.
\end{align}
\end{subequations}
Using the commutation relations of the fermionic modes and the fact that $\bar{\mathcal{A}}$ contains
only the bosonic modes, the tree-level graviton to graviton amplitude is given by
\beq
	\bra{f}\bar{\mathcal{A}}\ket{i}= \ep_{1\m \n}\:\ep_{2\r \s }\:\eta^{\m \r}\:\eta^{\n \s}
	\bra{0;0}\bar{\mathcal{A}}\ket{0;0}=\tr(\ep_1^T\:\ep_2)\: \mathcal{A}(s,\mathbf{b}).
\eeq
It is clear from the definition of $\mathcal{A}(s,\mathbf{b})$ that this will reproduce the expected
result in $q$-space upon a Fourier transformation.

Now we will show that the tree-level amplitude for a state of mass $\aph M_1^2=4n$ to become a state of 
mass $\aph M_2^2=4n'$ after scattering from a D-brane in the high energy limit is given by the corresponding
matrix element of the operator $\bar{\mathcal{A}}$ between the initial and final states. To do this
we must show that
\beq
	\bra{n'}\bar{\mathcal{A}}\ket{n}=\mathcal{A}_{n,n'}\left(s,\mathbf{b}\right)
\eeq
where $A_{n,n'}(s,\mathbf{b})$ is the Fourier transform of equation~(\ref{eq:result_mn}). We will use the standard 
oscillator modes representation of the initial and final states
\begin{subequations}
\begin{align}
	\ket{n}&= \varep_{\m_1\ldots \m_n \ah}\:\widetilde{\varep}_{\n_1\ldots \n_n \b}\:\frac{1}{n!}
		\prod_{i=1}^n\ah^{\m_i}_{-1}\widetilde{\ah}^{\n_i}_{-1}\:\p_{-\frac{1}{2}}^\ah \pt_{-\frac{1}{2}}^\b \ket{0;0},
	\\
	\ket{n'}&=\mathcal{G}_{\r_1\ldots\r_{n'}\s}\: \widetilde{\mathcal{G}}_{\l_1\ldots\l_{n'} \g}\:\frac{1}{n'!}
		\prod_{j=1}^{n'}\ah^{\r_j}_{-1}\widetilde{\ah}^{\l_j}_{-1} \p_{-\frac{1}{2}}^{\s} \pt_{-\frac{1}{2}}^{\g}\ket{0;0}.
\end{align}
\end{subequations}
Proceeding as before we obtain:
\beq
	\bra{n'}\bar{\mathcal{A}}\ket{n}=K_{\r_1\ldots\r_{n'} \l_1\ldots\l_{n'}\m_1\ldots \m_n\n_1\ldots \n_n}\bra{0;0}\prod_{i=1}^{n'}\ah^{\r_i}_1
	\widetilde{\ah}^{\l_i}_1 \bar{\mathcal{A}}\prod_{j=1}^n\ah^{\m_j}_{-1}\widetilde{\ah}^{\n_j}_{-1}\ket{0;0},
	\label{eq:a_mn}
\eeq
where the polarisations are contained within the tensor
\beq
	K_{\r_1\ldots\r_{n'} \l_1\ldots\l_{n'}\m_1\ldots \m_n\n_1\ldots \n_n}=\frac{1}{n!n'!}\mathcal{G}_{\r_1\ldots\r_{n'}\s}\eta^{\s \ah} 
	\varep_{\m_1\ldots \m_n \ah} \:	\widetilde{\mathcal{G}}_{\l_1\ldots\l_{n'} \g} \eta^{\g \b} \widetilde{\varep}_{\n_1\ldots \n_n \b}.
	\label{eq:big_k}
\eeq
To prove that the amplitudes derived in Section \ref{sec:ope_methods} are well described by the eikonal,
we compute the values of equation~(\ref{eq:a_mn}) and compare them with results stated in 
equations~(\ref{eq:polarisation}) and (\ref{eq:result_mn}).

In expanding the operator $\bar{\mathcal{A}}$ in oscillator modes it can be seen that very few terms will 
contribute a nonzero value to this matrix element. Since we are considering states belonging to the leading Regge trajectory, 
 these terms can only be composed of the modes 
$\ah_{\pm1}$, $\widetilde{\ah}_{\pm1}$, and since they must be normal ordered there can be at most $n$ 
occurrence of the operators $\ah_1$, $\widetilde{\ah}_1$ and $n'$ occurrences of the operators $\ah_{-1}$, 
$\widetilde{\ah}_{-1}$. As a result we only expect nonzero contributions from terms of order $2|n'-n|$ 
through to $2(n'+n)$. Furthermore, the terms in $\bar{\mathcal{A}}$ containing $k$ oscillators are 
generated by 
\beq
	\frac{1}{k!}\frac{\part^k \mathcal{A}(s,\mathbf{b})}{\part b^{\m_1}\ldots \part b^{\m_k}}\overline{X^{\m_1}\ldots X^{\m_k}},
\eeq
and the invariance of the partial derivative under a change in order of differentiation will result in the 
oscillators being symmetric under exchange of their indices, so rather than keeping track of these indices 
we need only count how many ways we can generate oscillator terms of the form given above. 

Using this one can deduce that we can substitute for $\bar{\mathcal{A}}$ in eq.~(\ref{eq:a_mn}) the following 
quantity
\begin{multline}
	\sum_{a,b=0}^{\min \{n,n'\}}\left(\frac{\aph}{2}\right)^{n+n'-a-b}\frac{(-1)^{n+n'}}{(n-a)!(n-b)!(n'-a)!(n'-b)!}
	\frac{\part^{2(n+n'-a-b)}\mathcal{A}(s,\mathbf{b})}{\part b^{i_1}\ldots \part b^{\ell_{n-b}}} \\
	\times \ah^{i_1}_{-1}\ldots \ah^{i_{n'-a}}_{-1}\widetilde{\ah}^{j_1}_{-1}\ldots \widetilde{\ah}^{j_{n'-b}}_{-1}\ah^{k_1}_1\ldots \ah^{k_{n-a}}_1
	\widetilde{\ah}_1^{\ell_1}\ldots \widetilde{\ah}^{\ell_{n-b}}_1.
\end{multline}
This has been written such that $a$ and $b$ will be seen to count the number of 
contractions between the polarisation tensors and they take on the range of values 
$a,b=0,1,\ldots,\min \{n,n'\}$; naturally we must count how many ways these terms may be generated using 
the symmetry of the partial derivatives in the expansion of $\bar{\mathcal{A}}$. The result of this substitution is
the amplitude shown below 
\begin{multline}
	\mathcal{A}_{n,n'}(s,\mathbf{b})=K_{\r_1\ldots\r_{n'} \l_1\ldots\l_{n'}\m_1\ldots \m_n\n_1\ldots \n_n}
	\sum_{a,b=0}^{\min \{n,n'\}}(-1)^{n+n'}\left(\dfrac{\aph}{2}\right)^{n+n'-a-b} C_{n,n'}(a)C_{n,n'}(b)
	\\ \times\frac{\part^{2(n+n'-a-b)}\mathcal{A}(s,\mathbf{b})}{\part b^{i_1}\ldots \part b^{\ell_{n-b}}} 
	\d^{\r_1 i_1}\ldots \d^{\r_{n'-a}i_{n'-a}}\d^{k_1\m_1}\ldots \d^{k_{n-a}\m_{n-a}}\d^{\r_{n'-a+1}\m_{n'-a+1}}\ldots \d^{\r_{n'} \m_n}\\
	\times \d^{\l_1 j_1}\ldots \d^{\l_{n'-b}j_{n'-b}}\d^{\ell_1\n_1}\ldots \d^{\ell_{n-b}\n_{n-b}}\d^{\l_{n'-b+1}\n_{n'-b+1}}\ldots \d^{\l_{n'} \n_n}.
\label{eq:impact_param_amp}
\end{multline}
To see that this is indeed equivalent to equation~(\ref{eq:result_mn}), one can perform a Fourier
transform on equation~(\ref{eq:impact_param_amp}) to arrive at the expected result
\beq
A_{n,n'}(s,\mathbf{q})=K_{n,n'}(\mathbf{q},\ep,G)A(s,\mathbf{q}),
\eeq
where now the kinematic function coincides with the one given in equation~(\ref{eq:polarisation}) and 
$A(s,\mathbf{q})$ is as written in (\ref{eq:stripped_amp}). In principle we already have a simple prescription
for using the eikonal operator that would allow us to compute the one-loop correction to the above expression,
this being given by the quadratic term arising from the exponential in (\ref{eq:eikonal}); if pomeron vertex
operator methods can be extended to generate such corrections as well, then it would be interesting to see whether the
equivalence between these two approaches continues to hold.  

The amplitude written in momentum space has a simple structure; it is neatly factored into
the graviton amplitude and some modifying kinematic function which is itself simply composed
of a sum over all the ways one may saturate various contractions of the tensor~(\ref{eq:big_k}) with the
momentum $\mathbf{q}$. The amplitude written in impact parameter space, as in (\ref{eq:impact_param_amp}), 
has an index structure that is much more complicated than its Fourier transform in momentum space. 
However, this intricate structure is greatly simplified in the limit of very large impact parameters;
in such a limit the function $\mathcal{A}$ takes the form
\beq
	\mathcal{A}(s,\mathbf{b})\sim s\sqrt{\pi}\frac{\G \left(\tfrac{6-p}{2}\right)}{\G \left(\tfrac{7-p}{2}\right)}
		\frac{R_p^{7-p}}{b^{6-p}}+i\frac{s\pi}{\G \left( \tfrac{7-p}{2}\right)}\sqrt{\frac{\pi\aph s}{\ln \aph s}}
		\left( \frac{R_p}{\sqrt{2\aph \ln \aph s}}\right)^{7-p}e^{-\tfrac{b^2}{2\aph \ln \aph s}},
\eeq
where to reflect the change to coordinate space we choose to express the normalisation of the amplitude 
in terms of the scale $R_p$. To be more specific we will examine impact parameters for which 
$b \gg R_p \gg \sqrt{2\aph \ln \aph s}$, that is, those much larger than both the effective string length 
and the characteristic size of the D$p$-branes; this being the case, we shall ignore the 
imaginary part of $\mathcal{A}(s,\mathbf{b})$ and focus on the real part. The result of these 
considerations is that the dominant contribution to $\mathcal{A}_{n,n'}$ comes from the term with the 
least number of derivatives in $b$. Without loss of generality let us assume that $n\le n'$ and denote the
difference between the two by $\D=n'-n$, then this term is given by
\begin{multline}
	\mathcal{A}_{n,n+\D}(s,\mathbf{b})\sim s\sqrt{\pi}\frac{\G \left(\tfrac{6-p}{2}\right)}{\G \left(\tfrac{7-p}{2}\right)}
		R_p^{7-p}K_{\r_1\ldots\r_{n+\D} \l_1\ldots\l_{n+\D}\m_1\ldots \m_n\n_1\ldots \n_n}
	\left(-\dfrac{\aph}{2}\right)^{\D}\left(\frac{(n+\D)!}{\D!}\right)^2 \\
	\times \frac{\part^{2\D}b^{-(6-p)}}{\part b^{i_1}\ldots \part b^{j_\D}}
	\d^{\r_1 i_1}\ldots \d^{\r_\D i_\D}\d^{\r_{\D+1}\m_{1}}\ldots \d^{\r_{n+\D} \m_n}
	 \d^{\l_1 j_1}\ldots \d^{\l_{\D}j_{\D}}\d^{\l_{\D+1}\n_{1}}\ldots \d^{\l_{n+\D} \n_n}.
\label{eq:large_b}
\end{multline}
The derivative above can be decomposed into a product consisting of an overall factor $\D!\, b^{-(6+2\D-p)}$
multiplied by some tensor which may be determined from the Gegenbauer polynomials in the following way.
The Gegenbauer polynomial $C^{(\l)}_m(x)$ is given in terms of the hypergeometric functions by
\beq
	C^{(\l)}_m(x)=\binom{m+2\l-1}{m}{}_2F_1(-m,m+2\l;l+\tfrac{1}{2};\tfrac{1}{2}(1-x));
\eeq
the tensor we are interested in is obtained by taking $C^{(6-p)/2}_{2\D}(x)$ and 
for each term in this polynomial we can attach the appropriate index structure by substituting $b_{i_r}/|b|$ 
for each factor of $x$, pairing the remaining indices up with Kronecker deltas and symmetrising the result.
Since it will not be required for this analysis we shall make no attempt to write the generic form for this 
tensor and we shall instead write it as $K(\mathbf{b}, \ep,G)/(n!(n+\D)!)$ after taking the contractions 
with $K_{\r_1\ldots\r_{n'} \l_1\ldots\l_{n'}\m_1\ldots \m_n\n_1\ldots \n_n}$, this function can be thought 
of as a hyperspherical harmonic. Thus equation~(\ref{eq:large_b}) becomes
\beq
	\mathcal{A}_{n,n+\D}(s,\mathbf{b})\sim s\sqrt{\pi}\frac{\G \left(\tfrac{6-p}{2}\right)}{\G \left(\tfrac{7-p}{2}\right)}
		R_p^{7-p}K(\mathbf{b}, \ep,G)\left(-\dfrac{\aph}{2}\right)^{\D}\frac{(n+\D)!}{n! \D!}\frac{1}{b^{6+2\D-p}}.
\eeq
From this expression it is seen that these kinds of inelastic excitations of the string begin to contribute significantly to 
scattering processes for impact parameters $b\le b_\D$ where
\beq
	b_\D^{6+2\D-p}=\sqrt{\pi s}\frac{\G \left(\tfrac{6-p}{2}\right)}{\G \left(\tfrac{7-p}{2}\right)}
		R_p^{7-p}\left(\dfrac{\aph}{2}\right)^\D \frac{(n+\D)!}{n! \D!}.
\eeq
The indication is that if one first considers scattering at impact parameters so large that only elastic 
scattering  is relevant and then begins to reduce this impact parameter then the elastic channel will 
be gradually absorbed, first by small string transitions and by larger transitions as $b$ is further
decreased in size. However, care should be taken in drawing quantitive conclusions from the above result since
it examines only a single particular inelastic channel, whereas any physical process would have many others 
which aren't considered by this analysis. These issues are discussed in more depth in terms of the full S-matrix
in \citep{d'appollonio10}.
%
\section{Conclusions}
\label{sec:conclusion}
%
What we have seen in this work is that at high energies the small-angle scattering of bosonic string states 
on the leading Regge trajectory from a stack of $N$ D$p$-branes exhibits universal tree-level behaviour, 
provided the masses of these states remain finite. In equation~(\ref{eq:result_mn}) this behaviour is 
characterised by the dependence on the square of the momentum flowing parallel to the D-branes, $s$, which 
is contained entirely within the factor $(\aph s)^{\aph \tfrac{t}{4}+1}$.  This indicates that this process is
dominated by the exchange of states from the leading Regge trajectory between the string and the D-branes.
The mass of the initial and final string states will determine the form of the kinematic function
$K_{n,n'}(q,\ep,G)$ and thereby influence the dependence of the amplitude upon the transferred momentum $q$.
We have also shown that the tree-level result~(\ref{eq:result_mn}) in the Regge regime can be reproduced 
by the linear term in the perturbative expansion of the eikonal operator $\hat{\delta}(s,\mathbf{b})$
determined in \citep{d'appollonio10}. Since our analysis was limited to tree-level, it would be interesting 
to see if the agreement persists up to one-loop calculations by comparing the quadratic terms from the 
eikonal operator with the inelastic annulus amplitudes.

The eikonal operator should provide a complete description of the
string-brane interaction at high energy. It can be expanded in a double power series of the ratios $R_p/b$
and $\sqrt{\aph}/b$, the classical and string corrections respectively. In \citep{d'appollonio10} the 
explicit form of the eikonal operator was determined to leading order in $R_p/b$ and to all orders in 
$\sqrt{\aph}/b$. As in \citep{amati87a,amati87b}, we have shown that the string corrections are neatly
taken into account by a simple shift in the impact parameter for the Fourier transform of the tree-level
string amplitude. From these string corrections we have seen evidence that for large impact parameters
longitudinal excitations of the string as a result of interactions with the D-brane are absent at the
leading order in energy.
The analysis of \citep{d'appollonio10} was extended to the next-to-leading order corrections in $R_p/b$ only for the elastic
scattering angle, but not for the inelastic excitations. 
It would be interesting to clarify
the full structure of the eikonal operator and to understand how the string corrections enter at the
next-to-leading order in $R_p/b$. Thus for this purpose it would also be useful to move one step
further in the eikonal expansion, computing the one-loop amplitude for two massive states. Furthermore, it would
be of interest to see whether a simple extension to the effective vertex methods employed here can be
determined which continues to yield the full high energy expression beyond tree-level.

In this paper we kept the external states fixed while taking the high energy limit and therefore we could
consistently neglect their masses by demanding that these states satisfy the relation $\aph s >>n,n'$.
Since the string spectrum contains states with arbitrarily large masses one could consider this limit
and allow the masses to become very large by relaxing this condition. As long as $|n-n'|$ remains much
smaller than $\aph s$ then the momentum  transferred can be kept finite and the amplitudes could still
demonstrate Regge behaviour at high energy. For larger differences between the mass levels we expect that
the inelastic amplitudes will decay exponentially with the energy, the behaviour typical of string scattering 
processes at fixed angle. It would be interesting to make these observations more precise and to perform
a detailed study of the amplitudes between states of the leading Regge trajectory in the limit of large 
masses for the external states, using where applicable both OPE methods \citep{ademollo89,ademollo90,brower06}
and saddle-point approximations \citep{gross87a,gross87b}.

\vspace{2mm}
\noindent {\large \textbf{Acknowledgements} }

\vspace{2mm}
We would like to thank Giuseppe D'Appollonio, Paolo Di Vecchia, Rodolfo Russo, David Turton and Gabriele 
Veneziano for fruitful discussions. WB is supported by an STFC studentship. CM is supported by INFN and 
during the preparation of this work was supported also by a scholarship from Regione Sardegna.

%

%
\providecommand{\href}[2]{#2}\begingroup\raggedright\endgroup

\begin{thebibliography}{10}

\bibitem{gross87a}
D.~J. Gross and P.~F. Mende, ``{The High-Energy Behavior of String Scattering
  Amplitudes},'' {\em Phys. Lett.} {\bf B197} (1987) 129.

\bibitem{gross87b}
D.~J. Gross and P.~F. Mende, ``{String Theory Beyond the Planck Scale},'' {\em
  Nucl. Phys.} {\bf B303} (1988) 407.

\bibitem{gross88}
D.~J. Gross, ``{High-Energy Symmetries of String Theory},'' {\em Phys. Rev.
  Lett.} {\bf 60} (1988)
1229.

\bibitem{amati87a}
D.~Amati, M.~Ciafaloni, and G.~Veneziano, ``{Superstring Collisions at
  Planckian Energies},'' {\em Phys.Lett.} {\bf B197} (1987) 81.

\bibitem{amati87b}
D.~Amati, M.~Ciafaloni, and G.~Veneziano, ``{Classical and Quantum Gravity
  Effects from Planckian Energy Superstring Collisions},'' {\em
  Int.J.Mod.Phys.} {\bf A3} (1988) 1615--1661.

\bibitem{amati88}
D.~Amati, M.~Ciafaloni, and G.~Veneziano, ``{Can Space-Time Be Probed Below the
  String Size?},'' {\em Phys.Lett.} {\bf B216} (1989) 41.

\bibitem{mende89}
P.~F. Mende and H.~Ooguri, ``{Borel summation of string theory for Planck scale
  scattering},'' {\em Nucl.Phys.} {\bf B339} (1990) 641--662.

\bibitem{veneziano04}
G.~Veneziano, ``{String-theoretic unitary S-matrix at the threshold of
  black-hole production},'' {\em JHEP} {\bf 0411} (2004) 001,
  \href{http://arXiv.org/abs/hep-th/0410166}{{\tt hep-th/0410166}}.

\bibitem{amati07}
D.~Amati, M.~Ciafaloni, and G.~Veneziano, ``{Towards an S-matrix description of
  gravitational collapse},'' {\em JHEP} {\bf 0802} (2008) 049,
  \href{http://arXiv.org/abs/0712.1209}{{\tt 0712.1209}}.

\bibitem{d'appollonio10}
G.~D'Appollonio, P.~Di~Vecchia, R.~Russo, and G.~Veneziano, ``{High-energy
  string-brane scattering: Leading eikonal and beyond},''
  \href{http://arXiv.org/abs/1008.4773}{{\tt 1008.4773}}.

\bibitem{ademollo89}
M.~Ademollo, A.~Bellini, and M.~Ciafaloni, ``{Superstring regge amplitudes and
  emission vertices},'' {\em Phys. Lett.} {\bf B223} (1989)
318--324.

\bibitem{ademollo90}
M.~Ademollo, A.~Bellini, and M.~Ciafaloni, ``{Superstring regge amplitudes and
  graviton radiation at planckian energies},'' {\em Nucl. Phys.} {\bf B338}
  (1990)
114--142.

\bibitem{brower06}
R.~C. Brower, J.~Polchinski, M.~J. Strassler, and C.-I. Tan, ``{The Pomeron and
  gauge/string duality},'' {\em JHEP} {\bf 0712} (2007) 005,
  \href{http://arXiv.org/abs/hep-th/0603115}{{\tt hep-th/0603115}}.

\bibitem{fotopoulos10}
A.~Fotopoulos and N.~Prezas, ``{Pomerons and BCFW recursion relations for
  strings on D-branes},'' {\em Nucl.Phys.} {\bf B845} (2011) 340--380,
  \href{http://arXiv.org/abs/1009.3903}{{\tt 1009.3903}}.

\bibitem{gross89}
D.~J. Gross and J.~L. Manes, ``{The High-Energy Behaviour of Open String
  Scattering},'' {\em Nucl. Phys.} {\bf B326} (1989) 73.

\bibitem{polchinski98a}
J.~Polchinski, ``{String theory. Vol. 1: An introduction to the bosonic
  string},''. Cambridge, UK: Univ. Pr. (1998) 402 p.

\bibitem{polchinski98b}
J.~Polchinski, ``{String theory. Vol. 2: Superstring theory and beyond},''.
  Cambridge, UK: Univ. Pr. (1998) 531 p.

\bibitem{bianchi10}
M.~Bianchi, L.~Lopez, and R.~Richter, ``{On stable higher spin states in
  Heterotic String Theories},'' \href{http://arXiv.org/abs/1010.1177}{{\tt
  1010.1177}}.

\bibitem{green87a}
M.~B. Green, J.~H. Schwarz, and E.~Witten, ``{Superstring Theory. Vol. 1:
  Introduction},''. Cambridge, UK: Univ. Pr. ( 1987) 469 P. (Cambridge
  Monographs On Mathematical Physics).

\bibitem{garousi96}
M.~R. Garousi and R.~C. Myers, ``{Superstring Scattering from D-Branes},'' {\em
  Nucl. Phys.} {\bf B475} (1996) 193--224,
\href{http://arXiv.org/abs/hep-th/9603194}{{\tt hep-th/9603194}}.

\bibitem{callan86}
J.~Callan, Curtis~G., C.~Lovelace, C.~Nappi, and S.~Yost, ``{String Loop
  Corrections to beta Functions},'' {\em Nucl.Phys.} {\bf B288} (1987) 525.

\bibitem{divecchia99}
P.~Di~Vecchia and A.~Liccardo, ``{D branes in string theory. I},'' {\em NATO
  Adv. Study Inst. Ser. C. Math. Phys. Sci.} {\bf 556} (2000) 1--59,
\href{http://arXiv.org/abs/hep-th/9912161}{{\tt hep-th/9912161}}.

\bibitem{duo07}
D.~Duo, R.~Russo, and S.~Sciuto, ``{New twist field couplings from the
  partition function for multiply wrapped D-branes},'' {\em JHEP} {\bf 12}
  (2007) 042,
\href{http://arXiv.org/abs/0709.1805}{{\tt 0709.1805}}.

\bibitem{black10}
W.~Black, R.~Russo, and D.~Turton, ``{The Supergravity fields for a D-brane
  with a travelling wave from string amplitudes},'' {\em Phys.Lett.} {\bf B694}
  (2010) 246--251, \href{http://arXiv.org/abs/1007.2856}{{\tt 1007.2856}}.

\end{thebibliography}
\end{document}